\documentclass[preprint,nocopyrightspace]{sigplanconf}

\usepackage{amsmath}
\usepackage{url}
\usepackage{listings}
\usepackage{balance}

\begin{document}

\title{Vector Clocks in Coq: An Experience Report}

\authorinfo{Christopher Meiklejohn}
           {Basho Technologies, Inc.}
           {cmeiklejohn@basho.com}
           
\maketitle

\begin{abstract}
This report documents the process of implementing vector clocks in the Coq proof assistant for extraction and use in the distributed Dynamo-inspired data store, Riak.  In this report, we focus on the technical challenges of using Core Erlang code extracted from the proof assistant in a production-grade Erlang application, as opposed to verification of the model itself.
\end{abstract}
    
\category{D.1.3}{Programming Techniques}{Concurrent Programming}
\category{D.3.3}{Programming Techniques}{Language Constructs and
Features - abstract data types, patterns, control structures}

\terms{Algorithms, Design, Reliability, Formal Verification}

\keywords{Dynamo, Erlang, Riak, vector clocks}

\section{Introduction}
\label{sec:introduction}
In a distributed system, where data structures are replicated or shared between multiple communicating nodes, it is highly desirable to have a method for asserting that certain invariants are preserved after objects are manipulated.  In industry, one of the major approaches is using property-based testing utilities such as QuickCheck \cite{Arts:2008:TED:1411273.1411275} to verify that certain properties hold true over multiple inter-leavings of operations applied to these data structures.  However, given the amount of possible executions as the number of nodes or processes grow, exhausting that state space becomes more difficult.  An alternative approach for this type of verification is to use an interactive theorem prover. 

We explore vvclocks \cite{vvclocks}, an Erlang library that aims to provide a formally verified implementation of vector clocks for use in Erlang applications.  We describe the process of building this implementation, as described in 1988 \cite{fidge1988timestamps}, in the Coq \cite{coq} proof assistant and walk through the process of using Coq's code extraction utilities to generate Core Erlang \cite{coreerlang} code.
    
vvclocks leverages the $verlang$ project \cite{verlang}, an extension to Coq providing extraction of Core Erlang, written by Tim Carstens.  Given the differences between the source and target languages, we outline the process of adapting both the generated Core Erlang and Coq source to properly execute and compile in the Erlang environment.

To verify the applicability of this implementation, we replace the vector clock implementation in the open source Dynamo-based data store, Riak. We explore the process of writing an adapter to translate between the Coq-derived data structures and the base Erlang data structures, adapting the existing test suite to pass, as well as running the Riak data store with the newly extracted implementation of vector clocks.

Based on our experience, we have identified a series of issues that makes this approach not immediately viable for use in production systems.  The issues are the following:

\begin{itemize}
\item Authoring code in a statically-typed language and performing extraction to a dynamically-typed language is problematic due to the required encoding of type information and constructors in the exported objects.
\item Required use of a subset of the source language due to inherent differences in the implementation of the destination language; for example, extracting functions that rely on currying to a language that does not support currying.
\item Required use of an adapter layer to perform translations between the native data structures of the destination language and the exported data structures of the source language.
\end{itemize}

The main contributions of this work are the following:

\begin{itemize}
\item A Coq module providing an implementation of vector clocks.
\item An extracted Erlang module providing an implementation of vector clocks, authored in Core Erlang, extracted from the Coq proof assistant.
\item An Erlang module used as glue code that provides helpers to perform type conversion between native Erlang types and Coq's data structures.
\item Details of where the extraction process was difficult; e.g. manual modifications to the extracted source or non-idiomatic Coq implementations.
\end{itemize}

The following goals are out of scope:

\begin{itemize}
\item We do not explore verification of the vector clock model, specifically because the proofs are tied very closely to the implementation of data structures that have changed extensively during the development of this research.
\item We do not focus on using the most efficient implementation of the data structures.  We have found that a simple representation although inefficient, is much easier to debug.
\end{itemize}

\section{Background}
\label{sec:background}
The verified vector clock implementation discussed below is authored in the Coq \cite{coq} proof assistant.  Coq's code extraction utilities are used to generate Core Erlang \cite{coreerlang} code, using a supporting library called $verlang$ \cite{verlang}.  This Core Erlang code is then compiled and used to replace one of the modules in the Erlang-based data store Riak.  

The following subsections provide background information about Coq, Core Erlang, $verlang$, and vector clocks.
 
\subsection{Coq}
\label{sec:background:coq}
Coq is an interactive theorem prover that implements a dependently-typed functional programming language called Gallina.  In addition, Coq provides a vernacular for stating function definitions and theorems and a tactic language for allowing the user to define custom proof methods.  Coq also provides the ability to perform extraction of certified programs as Scheme, Haskell or OCaml code through this vernacular. Figure~\ref{fig:ble_nat} provides an example of a fixed point computation authored in Coq.

\begin{figure}
\begin{lstlisting}[]
(** Less than or equal to comparson for natural numbers. *)
Fixpoint ble_nat (n m : nat) {struct n} : bool :=
  match n with
  | O => true
  | S n' =>
      match m with
      | O => false
      | S m' => ble_nat n' m'
      end
  end.
\end{lstlisting}
\caption{Fixpoint that computes less than or equal to over two natural numbers.  Example taken from Benjamin C. Pierce's "Software Foundations". \cite{sf}}
\label{fig:ble_nat}
\end{figure}

\subsection{Core Erlang}
\label{sec:background:coreerlang}
Core Erlang \cite{it:2000-030} is an intermediate representation of the Erlang programming language, designed to be operable on by other programs.  The major design goals of Core Erlang are providing a simple grammar that can be converted to normal Erlang programs, as well as providing a regular, concise structure to allow for code-walking tools to operate on programs translated to Core Erlang.  While the semantics of Core Erlang are not the focus of this paper, there is a sample Core Erlang function in Figure~\ref{fig:ble_nat_2}.

\begin{figure}
\begin{lstlisting}[]
'ble_nat'/2 = fun (_n, _m) -> 
  case _n of
    'O' when 'true' ->
        'True'
    {'S', _n@} when 'true' ->
        case _m of
          'O' when 'true' ->
              'False'
          {'S', _m@} when 'true' ->
              call 'vvclock':'ble_nat'
                   ( _n@
                   , _m@
                   )
         end
   end
\end{lstlisting}
\caption{Generated Core Erlang function that computes less than or equal to over two natural numbers, modeled using Peano numbers.  Note the recursive call to itself at line 10.}
\label{fig:ble_nat_2}
\end{figure}

\subsection{$verlang$}
\label{sec:background:verlang}
Verlang \cite{verlang} is an experimental project by Tim Carstens that adds the Core Erlang extraction target to Coq, in an attempt to enable formally verified Erlang development with the Coq proof assistant.  $verlang$ provides a mapping between MiniML, the language that serves as an intermediary translation step during extraction, to Core Erlang.

There are a number of interesting problems to address during the translation, which is outlined by Tim Carstens in the $verlang$ code repository. \cite{verlang}  

In summary:

\begin{itemize}
\item While Coq supports module nesting, there is no such concept in Core Erlang.  In the extracted code, nested modules are supported by name mangling.
\item Currying is not supported in Core Erlang and any code that relies on this produces code that does not execute correctly during extraction.
\item While Core Erlang differentiates between inter- and intra- module calls, the extraction code treats all calls to functions as inter-module calls, and fully qualifies them with the module name.
\item There is currently no known good way to handle the Erlang $receive$ primitive.
\end{itemize}

\subsection{Vector Clocks}
\label{sec:background:vector_clocks}
Vector clocks \cite{Lamport:1978:TCO:359545.359563} provide a mechanism for reasoning about the relationship of events in a distributed system, specifically focusing on differentiating between causal and concurrent events.

Vector clocks are traditionally modeled as a list of pairs, each pair being composed of an actor and an operation count.  As events occur in the system, the entire vector clock is shipped with the event.  Each actor in the event increments its own count.  This allows for a partial ordering to be established over events in the system by comparing each actor's count between two vector clocks.  We are able to reason about events that happen concurrently or causally by calculating whether the vector clocks descend or diverge from one another.

In the Riak system, objects are replicated across a series of nodes, and each object is stored with a vector clock representing each actor's modifications of the object.  When dealing with divergent values across replicas, vector clocks are used to reason about whether the value is a descendent of another, and effectively replaces it, or was written concurrently as another object, in that case both values need to be preserved.

\section{Implementation}
\label{sec:implementation}
The following subsections deal with the implementation of the vector clocks in Coq, the problems with the extracted Core Erlang code, and writing the adapter layer in Erlang.

\subsection{Vector clocks in Coq}
\label{sec:implementation:coq}
The package inside of Riak that provides the Erlang vector clock implementation is $riak\_core$.  In implementing vector clocks in Coq for use in Riak, we attempt to stick as closely as possible to the existing API.

For simplicity, we model each clock as a triple of natural numbers representing the actor, current count, and timestamp.  This can be seen in Figure~\ref{fig:types}.  The timestamps are used for pruning only, and are not critical to the semantics of the vector clock.  These timestamps are modeled as monotonically advancing natural numbers, similar to a UNIX timestamp.  We model vector clocks as a list of these clocks.

\begin{figure}
\begin{lstlisting}[]
(** Type definitions for actors, counts and timestamps. *)
Definition actor := nat.
Definition count := nat.
Definition timestamp := nat.

(** Model clocks as triples. *)
Definition clock := prod actor (prod count timestamp).

(** Model vector clocks as a list of clock triples. *)
Definition vclock := (list clock)%type.

(** Create a vector clocks. *)
Definition fresh : vclock := nil.\end{lstlisting}
\caption{Types specifications for vector clocks in Coq.}
\label{fig:types}
\end{figure}

The API we provide mimics the API exposed by $riak\_core$.  This includes the following functions: 

\begin{itemize}
\item $fresh$\\ Used to generate a new, empty vector clock
\item $increment$\\ Used to increment the vector clock for a particular actor.
\item $equal$\\ Used to test equality between two vector clocks.
\item $descends$\\ Used to determine if one vector clock is an ancestor of another.
\item $merge$\\ Used to merge two vector clocks.
\item $get\_counter$\\ Used to extract the count for a particular actor.
\item $get\_timestamp$\\ Used to extract the timestamp for a particular actor.
\item $all\_nodes$\\ Used to extract all actors that have incremented the vector clock.
\item $prune$\\ Given a series of timestamps representing the bounds for pruning, prune the vector clock and return the result of pruning.
\end{itemize}

Not included in this paper, but available in the provided source code, is the implementation of all of these functions.  Figure~\ref{fig:increment} is an example showing how the increment function destructures the vector clock, attempts to locate the appropriate actor if it exists, and either adds a new actor or increments an existing actors count.

There are a few important things to note about this example:

\begin{itemize}
\item We abstract the increment function that operates over timestamps.  This is important because Coq has no notion of generating an Erlang, or UNIX, timestamp.  We then patch the generated Core Erlang code to call timestamp aware replacements.
\item There are two anonymous functions passed as arguments here --- to the find and filter functions respectively ---  that in our initial version of the library were abstracted into their own function, as they are used by many of the API functions.  Because this relies on the use of partial application, these functions had to be inlined to ensure the extracted code compiled correctly.
\end{itemize}

\begin{figure}
\begin{lstlisting}[]
(** Increment a vector clock. *)
Definition increment (actor : actor) (vclock : vclock) :=
  match find (fun clock => match clock with 
                               | pair x _ => beq_nat actor x
                           end) vclock with
  | None => 
    cons (pair actor (pair init_count init_timestamp)) vclock
  | Some (pair x (pair count timestamp)) => 
    cons (pair x (pair (incr_count count) 
                       (incr_timestamp timestamp)))
         (filter (fun clock => match clock with 
                                 | pair x _ => 
                                   negb (beq_nat actor x)
                               end) vclock)
  end.
\end{lstlisting}
\caption{$increment$ function for a vector clock.}
\label{fig:increment}
\end{figure}

\subsection{Code extraction to Core Erlang}
\label{sec:implementation:coreerlang}
When performing the code extraction, we are required to extract not only the code of the vector clocks implementation itself, but also all of the supporting Coq libraries used in our implementation.  In our example, this includes modules supporting the core Coq datatypes: specifically Peano numbers, the List module, and a module providing arithmetic equality over natural numbers.

In extracting our vector clock library to Core Erlang, we encounter numerous problems with our implementation, but none with the extraction of the core libraries.  The following subsections provide details and an example of each of these issues, with the workaround we identified.

\subsubsection{Missing data constructors}
\label{sec:implementation:constructors}
The first problem we experience is that the data constructors do not export correctly.  Figure~\ref{fig:fresh} shows our addition to the code to resolve the problem of the missing data constructor, $fresh$.

\begin{figure}
\begin{lstlisting}[]
'fresh'/0 = fun () ->
  []
\end{lstlisting}
\caption{$fresh$ vector clock data constructor}
\label{fig:fresh}
\end{figure}

\subsubsection{Incorrectly qualified calls}
\label{sec:implementation:incorrect}
We also run into problems related to calls that are incorrectly qualified that have to be manually modified. 

Specifically, we run into cases where files containing modules are overqualified, with the filename repeated in the module name (as in Figure~\ref{fig:nested}) and cases where function calls are missing their arity, resulting in a failed function call at runtime (as in Figure~\ref{fig:arity}).

The workaround we identified for dealing with these issues was to manually modify the extracted code to fix the locations with missing arities, or incorrectly qualified functions.

\begin{figure}
\begin{lstlisting}[]
call 'vvclock.VVClock':'ble_nat'
     ( _actor
     , _a
     )
\end{lstlisting}
\caption{Incorrectly qualified call with module name.}
\label{fig:nested}
\end{figure}

\begin{figure}
\begin{lstlisting}[]
'descends'/2 = fun (_vc1, _vc2) ->
  case call 'Coq.Lists.List':'fold_left'
            ( 'descends@'
            , _vc2
            , { 'Pair'
              , 'True'
              , _vc1
              }
            ) of
\end{lstlisting}
\caption{Missing arity when passing function to the fold\_left call next to $'descends@'$.}
\label{fig:arity}
\end{figure}

\subsubsection{Lack of currying}
\label{sec:implementation:currying}
Figure~\ref{fig:currying} provides an example of using currying in Coq.  In this example, the $find''$ method returns a closure that returns true if a given actor is associated with this clock.  

In this example, the extracted Core Erlang code does execute, but has turned a arity-1 call that returned a function into a arity-2 call that immediately computes a final result.

The workaround we identified for this issue was to manually inline all of these function definitions, which, in addition to being non-idiomatic, leads to a large amount of code duplication in the implementation.

\begin{figure}
\begin{lstlisting}
(** Source Coq function definition. *)
Definition find'' (actor : actor) :=
  fun clock : clock => match clock with
                           | pair x _ => negb (beq_nat actor x)
                       end.
                    
%% Generated Core Erlang function.
'find@'/2 = fun (_actor, _clock) -> 
  case _clock of
    { 'Pair'
    , _c
    , _x
    } when 'true' ->
        call 'Coq.Arith.EqNat':'beq_nat'
             ( _actor
             , _c
             )
   end
\end{lstlisting}
\caption{Coq code that generates incorrect Erlang code when extracted.}
\label{fig:currying}
\end{figure}

\subsection{Erlang adapter between Riak and vvclocks}
Next, we look at the adapter layer that is required to convert between the Coq-derived data structures exported to Core Erlang and the data structures provided by Erlang.  The following subsections look at the conversion of Peano numbers, boolean types, UNIX timestamps, and data located in the application environment. 

\subsubsection{Type conversion}
$verlang$ models the abstract data types provided by Coq using $n$-sized tuples with the first position of the tuple representing the constructor name, and the remaining $n-1$ slots of the tuple as the constructor arguments.  Nested function applications, as used in recursive or inductively defined data types, are modeled as nested tuples.  We see two examples of this below, specifically when working with booleans and natural numbers. 

\begin{figure}
\begin{lstlisting}[]
%% @doc Convert a natural number into a peano.
natural_to_peano(0) ->
    'O';
natural_to_peano(Natural) ->
    {'S', natural_to_peano(Natural - 1)}.

%% @doc Convert a peano number into a natural.
peano_to_natural('O') ->
    0;
peano_to_natural({'S', Peano}) ->
    1 + Peano.
\end{lstlisting}
\caption{Conversion of Peano numbers to Erlang's built in numerics.}
\label{fig:peano_conversion}
\end{figure}

First, we look at the modeling of naturals in Coq, using Peano numbers as provided by $Coq.Init.Peano$.  Concretely, these numbers are modeled using an inductive data type with two constructors: one for the base case that takes no arguments $O$, and one for the inductive case that takes an argument of a Peano number, $S$.  For example, to represent the Arabic numeral $2$ the following would be used: $S\ S\ O$.  This is translated to the following Erlang structure: $\{'S', \{'S', 'O'\}\}$.  Figure~\ref{fig:peano_conversion} shows our wrapper function used to convert back and forth between these data types.

\begin{figure}
\begin{lstlisting}[]
%% @doc Compare equality of two vclocks.
equal(VClock1, VClock2) ->
    case vvclock:equal(VClock1, VClock2) of
        'True' ->
            true;
        'False' ->
            false
    end.
    
%% @doc Determine if one vector clock is an ancestor of another.
descends(VClock1, VClock2) ->
    case vvclock:descends(VClock1, VClock2) of
        'True' ->
            true;
        'False' ->
            false
    end.
\end{lstlisting}
\caption{Wrapper functions for converting between Coq's boolean type and Erlang's representation of booleans as atoms when dealing with equalities or inequalities.}
\label{fig:boolean_conversion}
\end{figure}

Next, we look at booleans which are implemented as part of the $Coq.Bool.Bool$ module.  In Coq, booleans are modeled as an abstract data type with two constructors $True$ and $False$.  In Erlang, booleans are modeled using atoms, specifically the $true$ and $false$ atom, which are immutable constants.  We write a series of small wrapper function, as seen in Figure~\ref{fig:boolean_conversion}, which call into our exported module, and pattern match on the return value.

\subsubsection{Timestamps}
Timestamps are another area where we have to provide wrapper functions for performing type conversions.  As Coq has no notion of what a UNIX timestamp is, our original implementation of the $prune$ method models timestamps as monotonically advancing natural numbers.

To account for this, we provide helper functions that are used to convert Erlang's notion of a timestamp to a Peano number that can be used by the functions exported by Coq.  Figure~\ref{fig:timestamps} provides an example of this conversion.

\begin{figure}
\begin{lstlisting}[]
%% @doc Return natural timestamp.
timestamp() ->
    calendar:datetime_to_gregorian_seconds(erlang:universaltime()).

%% @doc Peanoized timestamp.
peano_timestamp() ->
    term_to_peano(timestamp()).
\end{lstlisting}
\caption{Conversion of Erlang's representation of UNIX timestamps to Peano numbers.}
\label{fig:timestamps}
\end{figure}

\subsubsection{Actors}
The original $increment$ function arguments, as provided by $riak\_core$'s vector clock implementation, includes an actor ID, which is typically an atom but can be any Erlang term.  In our Coq implementation, we model this as string, which is an inductively defined data type over ASCII characters, which are themselves defined inductively over eight bits.  See Figure~\ref{fig:ascii} for an example.

\begin{figure}
\begin{lstlisting}[]
Definition zero := Ascii false false false false false false false false.
\end{lstlisting}
\caption{Coq representation of ASCII character.\cite{coq:ascii}}
\label{fig:ascii}
\end{figure}

\subsubsection{Environment variables}
The original $prune$ function, as provided by $riak\_core$'s vector clock implementation, takes three arguments: the vector clock, a timestamp, and an object that provides a series of settings which should apply to the vector clock, specifically referred to as $bucket\ properties$.  A simple dictionary-like data structure stores these properties.

Given that we cannot operate over this data structure in Coq, we break apart our $certified$ prune function as implemented in Coq from our $wrapper$ function, which extracts these values out of the environment and directly passes them as formal arguments.  Figure~\ref{fig:env} provides an example of this.

\begin{figure}
\begin{lstlisting}[]
%% @doc Prune vclocks.
prune(VClock, _Timestamp, BProps) ->
    Old =   term_to_peano(get_property(old_vclock, BProps)),
    Young = term_to_peano(get_property(young_vclock, BProps)),
    Large = term_to_peano(get_property(large_vclock, BProps)),
    Small = term_to_peano(get_property(small_vclock, BProps)),
    vvclock:prune(VClock, Small, Large, Young, Old).
\end{lstlisting}
\caption{Wrapper function for dealing with information that must come from other parts of the system.  In this example, $get\_property$ is retrieving information from a Riak bucket, but could easily also be accessing something stored in the runtime environment.  The function $term\_to\_peano$ is used to convert the Erlang timestamp format to a Peano number, leveraging the $datetime\_to\_gregorian\_seconds$ function.}
\label{fig:env}
\end{figure}

\section{Evaluation}
When evaluating the vvclock package with the Riak data store, we immediately ran into problems related to how the inductive data structures have been modeled in Erlang.  We will explore one particular example, related to timestamps.

For example, to get the current time in Erlang and convert it to a UNIX timestamp takes anywhere between 5 - 8 microseconds.  However, a timestamp for the year 2014 to a Peano number takes much longer given the nested tuple encoding and how memory is allocated on the stack for tuples.  In a matter of minutes we were able to exhaust just over 8.0 GB of memory in attempting to encode a timestamp like $1390525760$, before the process terminated because we ran out of virtual memory.

To allow us to continue testing the rest of the library, we modified the timestamp encoding function to store the timestamp as a much smaller value which was easier to encode.  Once completed, we were able to successfully validate the existing test suite against our exported module.  While we were able to successfully validate that our unit tests passed, we still cannot use the library in a production Riak instance because of the problems of data structure encoding.

In addition, even though the vvclock module itself may be verified, the vclock module, which provides the adapter layer between the vvclock module and the rest of the system, must still be fully tested to verify correct behavior.  This most likely would be done using Erlang's existing unit testing utility, eunit, or the aforementioned QuickCheck.

Finally, debugging of modules compiled from Core Erlang is still difficult.  Tooling such as Dialyzer \cite{Sagonas:2007:DDE:1273920.1273926} and Proper \cite{proper} all rely on type specifications in the form of annotations, as well as debugging symbols in the compiled byte code; neither are supported by Core Erlang.

\section{Future Work}
Based on our experience of implementing the vvclocks library, we have identified a series of work items that would improve the viability of this approach for use in production systems authored in Erlang.

\subsection{Bugs in $verlang$}
\label{sec:futurework:bugs}
It is clear that some of the issues we experienced during extraction of the vector clocks implementation are related to bugs in the $verlang$ extraction process.

Specifically, we refer to the following issues:

\begin{itemize}
\item Incorrectly qualified function calls. (Section~\ref{sec:implementation:incorrect})
\item Missing data constructors. (Section~\ref{sec:implementation:currying})
\end{itemize}

It appears that these bugs can be addressed in the extraction process.

\subsection{Other Applications}
The original motivation for this work was to take the partially verified implementations \cite{distributed-data-structures} of two convergent replicated data types, specifically the G-Counter and PN-Counter as described by Shapiro, et. al \cite{shapiro:inria-00555588} and extract them for use in the Riak system.  

In performing the extraction of this code, we ran into similar problems as discussed in \ref{sec:futurework:bugs}.  However, rather than these issues occurring in the vector clocks implementation we authored, they occurred in the extraction of support libraries for the built-in data structures provided by Coq.  This made working with and debugging this implementation more difficult, because of the lack of control over the internal implementation of Coq's data structures.

Beacuse of this, we decided to simplify the model and focus on the implementation of a vector clock library using the basic list data structures provided by Coq, which provides similar semantics to a G-Counter.

\subsection{Adapter Layer}
It is highly desirable to eliminate the adapter layer for several reasons:

\begin{itemize}
\item $Performance$ The process of converting Coq data structures into Erlang data structures requires the encoding of constructor calls, in order to retain typing information in a language without abstract data types.  The overhead of performing this conversion is high, especially when dealing with large or complex data structures such as timestamps.
\item $Testing$ Erlang functions must be made to perform conversions of data structures to and from their Coq representations.  This is problematic in that it requires an additional level of testing where bugs in the creation of objects can be introduced.
\end{itemize}

However, this is problematic.  Consider the case of naturals, modeled as Peano numbers: any axioms that are defined, or theorems proven about functions executing over these types is only guaranteed because of the way the data structure has been modeled.  Specifically with inductive types, certain functions are only guaranteed to terminate because of properties held by the inductive hypothesis.  In the case of naturals, if we modeled this using Erlang's built in integer type, none of the properties regarding subtraction would hold once we passed zero. 

Regardless, we feel it is worth exploring an alternate means of extraction that attempts to leverage the built-in Erlang types more efficiently; sacrificing some safety for performance.  For example, a function operating over a Coq string would be exported to Core Erlang code that operated over an Erlang binary instead of the tuple encoded Coq structure.

\subsection{QuickCheck Property Generation}
Finally, we feel that the ability to generate QuickCheck models instead of generating source code for the data structures might be a more viable approach.  Based on a series of of axioms defined regarding the Erlang run-time system, and given the theorems defined in the proof assistant, we could generate a series of properties to guide development of the data structures that preserve invariants.

\appendix
\section{Code Availability}
All of the code discussed is available on GitHub under the Apache 2.0 License at \url{http://github.com/cmeiklejohn/vvclocks}.

\acks
Thanks to Andy Gross, Andrew Stone, Ryan Zezeski, Scott Lystig Fritchie, and Reid Draper for providing feedback during the research and development of this work.

\bibliographystyle{abbrvnat}
\bibliography{vvclocks}
\end{document}